\documentclass[12pt]{article}
\usepackage{amssymb,amsmath}
\usepackage[noblocks]{authblk}
\usepackage[top=0.75in, bottom=0.75in, left=0.75in, right=0.75in, dvips]{geometry}
\usepackage{caption}
\begin{document}
\textwidth 10.0in 
\textheight 9.0in 
\topmargin -0.60in
\title{Light Front Quantization with the Light Cone Gauge}
\author[1,2]{D.G.C. McKeon}
\author[1] {Chenguang Zhao}
\affil[1] {Department of Applied Mathematics, The
University of Western Ontario, London, ON N6A 5B7, Canada} 
\affil[2] {Department of Mathematics and
Computer Science, Algoma University, Sault St.Marie, ON P6A
2G4, Canada}
\date{}
\maketitle
   
\maketitle 
\noindent
email: dgmckeo2@uwo.ca

\begin{abstract}
The Dirac procedure for dealing with constraints is applied to the quantization of gauge theories on the light front.  The light cone gauge is used in conjunction with the first class constraints that arise and the resulting Dirac brackets are found.  These gauge conditions are not used to eliminate degrees of freedom from the action prior to applying the Dirac constraint procedure.  This approach is illustrated by considering Yang-Mills theory and the superparticle in a $2 + 1$ dimensional target space.
\end{abstract}

\section{Introduction}

Ever since Dirac introduced the idea of light front quantization [1], this approach has received attention.  It is related to working in the infinite momentum frame [2], and has proved useful in such diverse areas as gauge theories [3-11], supersymmetry [12], general relativity [13-16] and superstrings [17].

Quite often, the light cone gauge is used to simply eliminate variables occurring in the original gauge invariant action and then the resulting reduced action is quantized on the light front.  However, if one were to follow the Dirac procedure for quantizing gauge systems [18-19], one should first identify and then classify all constraints in a system and then introduce a gauge condition to accompany each of the first class constraints.  This procedure can be applied when using light-front variables.  It need not result in the same quantized theory that arises if the light cone gauge is used at the outset to eliminate ``superfluous'' degrees of freedom before applying the Dirac procedure.  We will illustrate this by considering Yang-Mills theory and the superparticle.  In both of these examples, strict adherence to the Dirac procedure yields Dirac brackets which are different from what is obtained by using gauge conditions to eliminate degrees of freedom prior to involving Dirac's approach.  This would indicate that it would be in order to re-examine conclusions reached by applying gauge conditions to eliminate degrees of freedom in light-front quantization at the outset.

In an appendix, we show how the generator of a local gauge transformation can be found from first class constraints in a model in which primary second class constraints occur, thereby extending the discussion of ref. [23].

\section{Yang-Mills Theory and the Light-Cone}

When a covariant vector $a^\mu \quad(\mu = 0, 1, \ldots , D-1$ with $g_{\mu\nu} = \rm{diag} (+, - \ldots))$ has light front coordinates
\begin{align}
\begin{split}
a^{\pm} &= \frac{1}{\sqrt{2}} (a^0 \pm a^{D-1})\\
a^i &= a^\mu \quad(\mu = 1 \ldots D-2)
\end{split}
\end{align}
so that
\begin{equation}
a \cdot b = a^+ b^- + a^- b^+ - a^ib^i ,
\end{equation}
then the Yang-Mills (YM) action is
\begin{align}
S_{YM} &= \int d^dx \left( - \frac{1}{4} F_{\mu\nu}^a F^{a\mu\nu}\right)\nonumber \\
&= \int d^dx \left(\frac{1}{2} F^{a+-}F^{a+-} + F^{a+i} F^{a-i} - \frac{1}{4} F^{aij} F^{aij} \right)
\end{align}
where
\begin{equation}
F^{a\mu\nu} = \partial^{\mu} A^{a\nu} - \partial^\nu A^{a\mu} + \epsilon^{abc} A^{b\mu} A^{c\nu}.
\end{equation}
This action, as well as ones in which $A^{a\mu}$ is coupled with spinor and/or scalar fields, has been analyzed in a number of papers [3-11], most often by reducing the number of independent fields in the initial action through imposition of a gauge condition
\begin{equation}
A^{a+} = 0
\end{equation}
and using any resulting equation of motion that is independent of the ``time'' derivative
\begin{equation}
\partial^+ f \equiv \dot{f}.  
\end{equation}
We will instead apply the Dirac constraint formalism [18-19] to the action of eq. (3), imposing gauge conditions in conjunction with first class constraints that arise.  This has been done when applying path integral quantization to the action of eq. (3) [11]. Dirac's canonical procedure has been applied to the light-front formulation of the $U(1)$ limit of Yang-Mills theory in ref. [28]. In ref. [15, 16], this approach has been used to analyze the spin-two action (ie, linearized gravity).  We will also show how the first class constraints arising from the action of eq. (3) lead to a generator of the usual gauge transformation
\begin{align}
\delta A_\mu^a &= D_\mu^{ab} \theta^b\nonumber \\
& \equiv \left(\partial_\mu \delta^{ab} + \epsilon^{apb} A_\mu^p\right) \theta^b
\end{align}
despite the presence of second class constraints.

We begin by computing the canonical momenta 
\begin{subequations}
\begin{align}
\pi^a_i &= \partial \mathcal{L}_{YM}/\partial \dot{A}^{ai} = F^{a-i}\\
\pi^a_+ &= \partial \mathcal{L}_{YM}/\partial \dot{A}^{a+} = 0\\
\pi^a_- &= \partial \mathcal{L}_{YM}/\partial \dot{A}^{a-} = F^{a+-}.
\end{align}
\end{subequations}
Together, these result in the canonical Hamiltonian 
\begin{equation}
\mathcal{H}_c = \frac{1}{2}\pi_-^a \pi_-^a + \frac{1}{4} F^{aij} F^{aij} - A^{a+} \left( D^{abi} \pi_i^b + D^{ab-} \pi_i^b\right).
\end{equation}
Eq. (8a) is obviously a second class primary constraint
\begin{equation}
\theta_i^a = \pi_i^a - F^{a-i}.
\end{equation}
From the primary constraint of eq. (8b)
\begin{subequations}
\begin{align}
\phi_1^a &= \pi_+^a\\
\intertext {and the canonical Hamiltonian of eq. (9) we obtain the secondary constraint}
\phi^a_2 &= D^{abi} \pi_i^b + D^{ab-} \pi_-^b;
\end{align}
\end{subequations}
it is evident that $\phi_1^a$ and $\phi_2^a$ are both first class and that no further constraints arise.

The constraints of eqs. (9,11) have the Poisson bracket (PB) algebra 
\begin{subequations}
\begin{align}
\left\lbrace \phi_2^a, \phi_2^b\right\rbrace &= \epsilon^{abc}\phi_2^c\\
\left\lbrace \phi_2^a, \theta_i^b\right\rbrace &= \epsilon^{abc}\theta_i^c\\
\left\lbrace \theta_i^a(x), \theta_j^b(y)\right\rbrace &= -2\delta_{ij} D^{ab-}  \delta (x-y),
\end{align}
\end{subequations}
and so, by eq. (12c), we can eliminate the second class constraint $\theta_i^a$ by defining the Dirac bracket (DB)
\begin{equation}
\left\lbrace M, N\right\rbrace^* = \left\lbrace M,N \right\rbrace - \left\lbrace M, \theta_i^a (z) \right\rbrace \frac{-1}{2D_z^{ab-}} \delta(z-w) \left\lbrace \theta_i^b (w), N\right\rbrace .
\end{equation}

The non-trivial DB of eq. (12c) leads to a non-trivial contribution to the measure of the path integral if one were to use path integral quantization [11]. In the $U(1)$ limit considered in ref. [28], this contribution to the measure of the path integral becomes trivial.  We also note that the inverse operator $1/\partial^-$ arising in the $U(1)$ limit of eq. (13) is carefully defined in ref. [28].

As in eq. (A.7), we define the generator of the gauge transformation that leaves $S_{YM}$ of eq. (3) invariant to be
\begin{equation}
G = \mu_1^a \phi_1^a + \mu_2^a\phi_2^a
\end{equation}
with $\mu_1^a$ determined in terms of $\mu_2^a$ by those terms in eq. (A.11) at least linear in $\phi_2^a$,
\begin{equation}
\left( \dot\mu_1^a \phi_1^a + \dot{\mu}_2 \phi_2^a \right) +\left\lbrace \mu_1^a \phi_1^a + \mu_2^a \phi_2^a, \mathcal{H}_c \right\rbrace - \delta \mu_1^a \phi_1^a = 0
\end{equation}
which by eqs. (9, 12) leaves us with
\begin{equation}
G = \left( \dot{\mu}_2^a + \epsilon^{abc} A^{b+} \mu_2^c\right) \phi_1^a + \mu_2^a \phi_2^a .
\end{equation}
From eq. (16) we find the gauge transformation of eq. (7) with $\theta^a = \mu_2^a$, as expected.

As was done ref. [11], the first class constraints $\phi_I^a$ of eqs. (11a,b) are accompanied by gauge conditions $\gamma_I^a$ so that together $\phi_I^a$ and $\gamma_I^a$ form a set of second class constraints.  Here we will use the same gauge conditions that were suggested in ref. [11], and will proceed to find the resulting DB.

The constraint of eq. (11a) naturally leads to the gauge condition
\begin{equation}
\gamma_1^a = A^{a+}
\end{equation}
while that of eq. (11b) suggests
\begin{subequations}
\begin{align}
\gamma_{2I}^a &= A^{a-}\\
\intertext{or}
\gamma_{2II}^a &= \partial^i A^{ai}.
\end{align}
\end{subequations}
(The gauge conditions of eq. (18) are distinct from the gauge condition $\partial_\mu A^\mu = 0$ considered in ref. [28].)  Having already eliminated $\theta_i^a$ of eq. (10) by defining the DB of eq. (13), we can now eliminate $\phi_1^a$ and $\gamma_1^a$ by the ``second stage'' DB
\begin{equation}
\left\lbrace M,N\right\rbrace^{**} = \left\lbrace M,N\right\rbrace^{*} - 
\left[ \left\lbrace M, \pi_+^a(z)\right\rbrace^{*}\delta (z-w) \left\lbrace A^{a+} (w),N\right\rbrace^{*} - ( M \leftrightarrow N)\right].
\end{equation}
(Unlike ref. [28], we eliminate second class constraints in stages.) In the same way $\phi_2^a$ and $\gamma_{2I}^a$ give rise to a ``third stage'' DB.  This involves using constraints in stages
\begin{subequations}
\begin{align}
\left\lbrace \gamma_{2I}^a, \phi_2^b \right\rbrace^{**} &= - D^{ab-}_x \delta(x-y)\\
\left\lbrace \phi_2^a, \phi_2^b\right\rbrace^{**} &= \epsilon^{abc} \phi_2^c - \left[ \epsilon^{apm}\theta_i^m (x) \right] \frac{-1}{2D^{pq}_x} \delta (x-y) \left[ -\epsilon^{bqn} \theta_i^n (y)\right].
\end{align}
\end{subequations}
When forming the DB to eliminate $\gamma_{2I}^a$ and $\phi_2^a$, one can set $\phi_2^a$ and $\theta_i^a$ to zero in eq. (20b) and so our third stage DB is 
\begin{align}
\left\lbrace M,N\right\rbrace^{***}& = \left\lbrace M,N\right\rbrace^{**} - \left[ \left\lbrace M, \phi_2^a (z)\right\rbrace^{**} \frac{-1}{D_z^{ab-}} \delta(z-w)\right. \nonumber \\
& \left. \left\lbrace \gamma_{2I}^b(w),N\right\rbrace^{**} - (M \rightleftharpoons N) \right].
\end{align}
Computing the third stage DB when using the gauge condition $\gamma_{2II}^a$ of eq. (18b) in conjunction with the first class constraint $\phi_2^a$ of eq. (11b) is more involved.  Eq. (20b) still holds, but now we also have
\begin{equation}
\left\lbrace \gamma_{2II}^a, \gamma_{2II}^b \right\rbrace^{**} = \frac{1}{2} \partial^k \frac{1}{D^{ab-}_x} \partial^k \delta (x-y)
\end{equation}
as well as
\begin{equation}
\left\lbrace \gamma_{2II}^a, \phi_2^b \right\rbrace^{**} = -\partial^i D^{abi} \delta (x-y) - \frac{1}{2} \partial^i  \frac{1}{D^{aq-}_x} \delta(x-y)\epsilon^{bqr} \theta_i^r(y).
\end{equation}
Again, in eqs. (20b, 22, 23) we can set $\phi_2^a = \theta_i^a = 0$ when forming the DB to eliminate $\gamma_{2II}^a$ and $\phi_2^a$; since 
\begin{equation}
\left(
\begin{array}{cc}
 \frac{1}{2} \partial ^k \frac{1}{D^{ab-}} \partial^k & -\partial^i D^{abi}\\
- D^{abi} \partial^i & 0
\end{array}\right) =
\left(
\begin{array}{cc}
0 & \frac{-1}{D^{ab-}\partial^i}\\
\frac{-1}{\partial^iD^{abi}} &\quad - \frac{1}{2} \frac{1}{\partial^iD^{api}} \partial^k \frac{1}{D^{pq-}} \partial^k \frac{1}{D^{qbj}\partial^j} 
\end{array}
\right)
\end{equation}
we find that
\begin{align}
\left\lbrace M, N \right\rbrace^{***}& = \left\lbrace M, N \right\rbrace^{**} - \bigg[ 
\left\lbrace M,\gamma_{2II}^a (z)  \right\rbrace^{**} \frac{-1}{\partial^j D^{abj}_z} \delta(z-w) \nonumber \\
& \quad \left\lbrace \phi_2^b(w), N\right\rbrace^{**} - (M \rightleftharpoons )N \bigg]\nonumber\\
& \quad - \bigg[ \left\lbrace M, \phi_2^a (z)\right\rbrace^{**} \left( \frac{-1}{2}\right)\frac{1}{\partial^iD^{api}_z}\partial^k \frac{1}{D^{pq-}_z} \partial^k \frac{1}{D^{qbj}_z\partial^j}\delta(z-w)\nonumber \\
 &\qquad \left\lbrace \phi_2^b (w), N\right\rbrace^{**}\bigg].
\end{align}
For example, from eq. (25) it follows that we have the novel DB
\begin{align}
&\left\lbrace A^{ai}(x),  A^{bj}(y)\right\rbrace^{***} \nonumber \\
&\quad = \frac{1}{2} \bigg[ -\delta^{ij} \frac{1}{D^{ab-}}  + \frac{1}{D^{ap-}} \partial^i \frac{1}{D^{pqk}\partial^k} D^{qbj} + D^{api} \frac{1}{\partial^k D^{pqk}}\partial^j 
\frac{1}{D^{qb-}}\nonumber \\
&\quad - D^{api} \frac{1}{\partial^kD^{pqk}}\partial^m \frac{1}{D^{qr-}} \partial^m 
\frac{1}{D^{rs\ell}\partial^\ell} D^{sbj}\bigg]\delta(x-y).
\end{align}
We see from eq. (26) that
\begin{equation}
\left\lbrace \partial^i A^{ai}, A^{bj}\right\rbrace^{***} = 0
\end{equation}
which is consistent with the gauge condition of eq. (18b).  In the $U(1)$ limit, eq. (26) reduces to
\begin{equation}
\left\lbrace A^i(x), A^j(y) \right\rbrace^{***} = \frac{1}{2}\left( -\delta^{ij} + \frac{\partial^i\partial^j}{\partial^k\partial^k} \right) \frac{1}{\partial^-}\delta(x-y).
\end{equation}

The form of eq. (26) ensures that only the transverse components of $A^{ai}$ contribute to the dynamics of Yang-Mills theory when using light-front quantization and the gauge condition of eq. (18b).  When working in $3 + 1$ space-time dimensions, we see that the only two physical degrees of freedom are $A^{a-}$ and the transverse components of $A^{ai}$ when using the gauge condition of eq. (18b); if the gauge condition of eq. (18a) were used, then all of the degrees of freedom reside in $A^{ai} \quad (i = 1,2)$.

We thus see that applying the Dirac canonical analysis to YM theory right from the outset (ie, only introducing constraints after the first class constraints which follow from the initial YM action when written in light front coordinates) yields different DB than what arises when the light cone gauge is used to eliminate degrees of freedom from the YM action before employing the Dirac formalism.

The spin two action has been treated in a manner consistent with the approach used here with YM theory in ref. [15].

We now turn to examining the superparticle in the light cone gauge.

\section{The Superparticle and the Light Cone}

The superparticle [20] has Bosonic variables $x^\mu(\tau)$ and Fermionic variables $\theta(\tau)$; its action is
\begin{equation}
S = \int d\tau \frac{1}{2e} \left( \dot{x}^\mu + i \dot{\overline{\theta}} \gamma^\mu \theta\right) \left( \dot{x}_\mu + i \dot{\overline{\theta}} \gamma_\mu \theta\right).
\end{equation}
A discussion of its constraint structure appears in ref. [21] (see also ref. [22]). In ref. [28], considering application of the Dirac formalism to systems involving Fermionic degrees of freedom was not yet feasible as degrees of freedom that were Grassmann degrees of freedom had not yet been introduced into the Dirac canonical formalism. Quite often, the light cone gauge conditions
\begin{subequations}
\begin{align}
x^+ &= p_+\tau\\
\gamma^+\theta &= 0
\end{align}
\end{subequations}
are used [17] to eliminate degrees of freedom from the action of eq. (29) prior to applying Dirac's formalism; below we will instead use the gauge conditions of eq. (30) in conjunction with the first class constraints arising from eq. (29).

The spinor $\theta$ has different properties in every dimension of the target space; we will restrict our attention to $2 + 1$ dimensions to simplify our discussion.  The conventions of ref. [21] will be used, so that 
\begin{align}
\gamma^0 = \sigma_2 \quad \gamma^1 = i\sigma_3 \quad \gamma^2 = i\sigma_1\\
\gamma^\mu \gamma^\nu = \eta^{\mu\nu} + i\epsilon^{\mu\nu\lambda}\gamma_\lambda\nonumber
\end{align}
\begin{align}
C &= -\gamma^0\\
\theta = C \overline{\theta}^T &= (-\gamma^0)(\theta^\dagger\gamma^0)^T\nonumber
\end{align}
so that
\begin{equation}
\theta = \left( \begin{array}{c}
u \\ d
\end{array} \right) =
\left( \begin{array}{c}
u^* \\ d^*
\end{array} \right).
\end{equation}
With this, we find that eq. (29) becomes
\begin{equation}
S = \int \frac{d\tau}{2e}\left[ \left( \dot{x}^0 + i (\dot{u} u + \dot{d}d)\right)^2 -
\left( \dot{x}^1 - i (\dot{u} d + \dot{d}u)\right)^2 - 
\left( \dot{x}^2 + i (\dot{u} u - \dot{d}d)\right)^2 \right]
\end{equation}
so that the momenta conjugate to $e$, $x^\mu$, $u$ and $d$ are
\begin{subequations}
\begin{align}
p_e &= 0\\
p_\mu &= \frac{1}{e} \left( \dot{x}^0 + i (\dot{u} u + \dot{d}d),
-\dot{x}^2 + i (\dot{u} d + \dot{d}u),
 -\dot{x}^2 - i (\dot{u} u - \dot{d}d)\right)\\
\pi_u &= -id p_1 + iu p_+\\
\pi_d &= i(dp_- - up_1)
\end{align}
\end{subequations}
where $p_{\pm} \equiv p_0 \pm p_2$.  We see that eqs. (34a,c,d) are primary constraints.  Following ref. [21], we treat $\sigma_1 = \pi_u + idp_1 - iup_+$ as a second class constraint and eliminate it by defining the DB
\begin{equation}
\left\lbrace M, N \right\rbrace^* = \left\lbrace M, N \right\rbrace - \left\lbrace M, \sigma_1 \right\rbrace \frac{1}{2ip_+} \left\lbrace \sigma_1, N \right\rbrace .
\end{equation}
With this DB, the constraint $\sigma_1 = \pi_d - idp_- + iup_1$ satisfies
\begin{equation}
\left\lbrace\sigma_2, \sigma_2 \right\rbrace^* = 2ip^2/p_+ .
\end{equation}
Since the canonical Hamiltonian is 
\begin{equation}
H_c = \frac{e}{2}p^2,
\end{equation}
we see that the primary constraint of eq. (35a) leads to the secondary first class constraint 
\begin{equation}
p^2 = 0,
\end{equation}
and hence by eq. (37), we see that once $\sigma_1$ has been taken to be second class, $\sigma_2$ becomes first class.  (The roles of $\sigma_1$ and $\sigma_2$ can be reversed.)

It is at this stage we introduce gauge conditions to accompany the first class constraints that have been derived.  In conjunction with 
\begin{equation}
\phi_1 = p_e, \quad \phi_2 = p^2, \quad \phi_3 = \sigma_2 \tag{40a,b,c}
\end{equation}
we introduce respectively
\begin{equation}
\gamma_1 = e-1, \quad \gamma_2 = x^+ - p_+\tau, \quad \gamma_3 = \gamma^+\theta = u .
\tag{41a,b,c}
\end{equation}
From the first class constraints of eq. (40), one can use the approach of ref. [23] to derive a generator of a set of Bosonic and Fermionic gauge transformations, the Fermionic ones being half of the so-called $\kappa$-symmetry transformations of ref. [24].  (The other half can be generated by reversing the rules of $\sigma_1$ and $\sigma_2$.)

Together, $\phi_I$ and $\gamma_I$ in eqs. (40, 41) constitute a set of second class constraints that can be eliminated by forming a ``second stage'' DB.  This involves inverting the matrix 
\begin{align}
M &= \left\lbrace (\gamma_1, \phi_1, \gamma_2, \phi_2, \gamma_3, \phi_3)^T, (\gamma_1, \phi_1, \gamma_2, \phi_2, \gamma_3, \phi_3) \right\rbrace^* \tag{42}\\
&= \left( \begin{array}{cccccc}
0 & 1 & 0 & 0 & 0 & 0 \\
-1 & 0 & 0 & 0 & 0 & 0 \\
0 & 0 & 0 & 2p_- & -u/p_+ & -2iup_1/p_+ \\
0 & 0 & -2p_- & 0 & 0 & 0 \\
0 & 0 & u/p_+ & 0 & i/2p_+ & -p_1/p_+ \\
0 & 0 & 2iup_1/p_+ & 0 & -p_1/p_+ & 2ip^2/p_+ \end{array}
\right).\nonumber
\end{align}
To find $M^{-1}$, we use the identity
\begin{equation}
\left(\begin{array}{cc} A & B \\
C & D \end{array} \right)^{-1} = 
\left(\begin{array}{cc} \Delta^{-1}  & -\Delta^{-1}BD^{-1} \\
-D^{-1}C\Delta^{-1} & D^{-1}+D^{-1}C\Delta^{-1}BD^{-1} \end{array} \right)
(\Delta = A - BD^{-1}C)\nonumber
\end{equation}
and $u^2 = 0$ (since u is Grassmann); we arrive at
\begin{equation}
M^{-1} =  \left( \begin{array}{cccccc}
0 & -1 & 0 & 0 & 0 & 0 \\
1 & 0 & 0 & 0 & 0 & 0 \\
0 & 0 & 0 & -1/2p_- & 0 & 0 \\
0 & 0 & 1/2p_- & 0 & -iu/p_- & 0 \\
0 & 0 & 0& -iu/p_+ & -2ip^2/p_- & -p_1/p_-\\
0 & 0 & 0 & 0 & -p_1/p_- & 1/2ip_- \end{array}
\right). \tag{43}
\end{equation}
From the resulting DB, it follows, for example that
\begin{align}
\left\lbrace x^1, x^2 \right\rbrace^{**} &= \left\lbrace x^1, x^2 \right\rbrace^* -
\left\lbrace x^1, \Phi^T \right\rbrace^* M^{-1} \left\lbrace \Phi , x^2 \right\rbrace\nonumber\\
&= \frac{p_1 \tau}{p_-} + \frac{iudp_0}{2p_+p_-} .\tag{44}
\end{align}
where $\Phi^T = (\gamma_1, \phi_1, \gamma_2, \phi_2, \gamma_3, \phi_3)^T$.  This result serves to illustrate how using the light cone gauge conditions of eq. (41) in conjunction with the first class constraints of eq. (40) (arrived at by applying Dirac's canonical procedure to the initial action of eq. (29)) leads to results different from those obtained by using eq. (41) to eliminate fields from eq. (29) and only then applying the Dirac procedure (as is normally done).

These considerations can also be applied to string theories.  For the Bosonic string, the action is [25]
\begin{equation}
S = \int d\tau d\sigma \left( \frac{1}{2} \sqrt{-g}\; g^{ab} x^A_{,a} x_{A,b} \right).\tag{45}
\end{equation}
The canonical momenta associated with $g^{ab}$ and $x_A$ are
\begin{subequations}
\begin{align}
I\!\!P_{ab} &= 0\tag{46a}\\
p_A &= \sqrt{-g} \left(g^{00} x_{A,0} + g^{01}x_{A,1}\right)\tag{46b}
\end{align}
\end{subequations}
which lead to the secondary first class constraints
\begin{subequations}
\begin{align}
\Sigma_S &= \frac{1}{2} \left(p^2 + x_{,1}^2\right)\tag{47a}\\
\Sigma_p &= p_{A} x_{,1}^A\tag{47b}
\end{align}
\end{subequations}
Both of these in principle should be accompanied by a suitable gauge condition.  However, the usual practice [29] is to use a single gauge condition (the ``light cone gauge'') and then using this to simplify the initial action of eq. (45).  Only at this stage is the Dirac procedure invoked.  A similar approach is generally used with the superstring.  (A discussion of the canonical structure of the superstring appears in ref. [26].)

\section{Discussion}

The Dirac procedure for treating the canonical structure of dynamical systems which have a local gauge invariance is well defined; all constraints are first obtained and then classified, and those which are first class are then paired with suitable gauge conditions.  All superfluous degrees of freedom arising on account of there being a local gauge symmetry are then eliminated by replacing the PB by a DB defined using both the first and second class constraints and the gauge conditions.  This procedure can be tedious, as can be seen by examining YM theory on the light front and the superparticle (as was done above).  Both of these systems can be simplified by using a ``light cone'' gauge condition to eliminate superfluous degrees of freedom at the outset from the classical action, and then using Dirac's procedure.  However, the DB one arrives at after using these two approaches need not be the same.  Being different would result in the two procedures leading to different quantum theories if one were to use the DB to define a quantum mechanical commutator.

When one employs path integral quantization, second class constraints can lead to highly non-trivial contributions to the measure of the path integral, as is discussed in ref. [20].  This is true when using the path integral to quantize the first order Einstein-Hilbert action [31] as well as when considering Yang-Mills theory on the light-front (see eq. [12c] and (43) above).

On the other hand, one often uses the Faddeev-Popov [32] approach to handling the path integral quantization of gauge theories.  This involves a ``factoring'' of the divergence arising in the path integral resulting from the presence of a gauge invariance.  In doing this, one (or more [33]) gauge conditions must be introduced, and ``ghost'' fields are used to cancel the contribution of non-physical degrees of freedom.  However, we note that there need not be agreement between results obtained from the path integral when the formalisms of refs. [30] and [32] are employed, especially if non-trivial second class constraints occur (such as in eq. [10] above).  Fortunately, in Yang-Mills theory the Faddeev-Popov approach to the path integral is equivalent to the canonical approach to the path integral when not using the light-front [34].

\section*{Acknowledgements}
Roger Macleod had useful advice.

\section*{Appendix}
In refs. [19, 23 ] it is shown how to obtain the generator of a gauge transformation for systems involving exclusively first class constraints.  Here we will extend this discussion to include the situation in which there are also primary second class constraints so that one can consider the light front formulation of Yang-Mills theory.

In the presence of primary second class constraints $\theta_\alpha$ and first class constraints $\phi_{A_{i}}$ (where $i$ denotes the generation of the constraint-primary is $i = 1$, secondary is $i = 2$ etc.), then suppose we have the PB algebra
\begin{equation}
\left\lbrace \theta_\alpha , \theta_\beta \right\rbrace = \Delta_{\alpha\beta},\tag{A.1}
\end{equation}
as well as 
\begin{equation}
\left\lbrace \phi_A , \phi_B \right\rbrace = C_{AB}^C \phi_C + C_{AB}^\alpha \theta_\alpha\tag{A.2}
\end{equation}
and
\begin{equation}
\left\lbrace \phi_A , \theta_\alpha \right\rbrace = C_{A\alpha}^\beta \theta_\beta +  + C_{A\alpha}^B \phi_B  .\tag{A.3}
\end{equation}
We then can deine the DB
\begin{equation}
\left\lbrace M, N \right\rbrace^* = \left\lbrace M,N \right\rbrace - \left\lbrace M, \theta_\alpha \right\rbrace \Delta_{\alpha\beta}^{-1} \left\lbrace \theta_\beta , N\right\rbrace . \tag{A.4}
\end{equation}
Upon using the constraints $\theta_\alpha$ and $\phi_{A_{i}}$, the canonical Hamiltonian $H_C$ can be defined 
\begin{equation}
H_C = p_i \dot{q}^i - L(q^i, \dot{q}^i); \tag{A.5}
\end{equation}
this leads to the extended Hamiltonian 
\begin{equation}
H_E = H_C + \sum_\alpha U_\alpha \theta_\alpha + \sum_{A_{i}} V_{A_{i}} \phi_{A_{i}} . \tag{A.6}
\end{equation}
If the sum over $A_i$ in eq. (A.6) is restricted to having $i = 1$ (ie, just the primary constraints) then $H_E$ reduces to $H_T$, the total Hamiltonian.

We now can consider the generator
\begin{equation}
G = \sum_{A_{i}} \mu_{A_{i}}\phi_{A_{i}}\tag{A.7}
\end{equation}
of ``gauge'' transformations that leave the extended action $S_E$ invariant, that is the change induced by $G$ on a dynamical quantity $f$ is given by
\begin{equation}
\delta f = \left\lbrace f, G \right\rbrace . \tag{A.8}
\end{equation}
The change in the extended action is given by
\begin{align}
\delta S_E &= \int dt \;\delta\left( p_i \dot{q}^i - H_E\right)\nonumber \\
&\int dt \bigg[ \delta p_i \dot{q}^i + p_i \delta \dot{q}_i - \left\lbrace H_C, G \right\rbrace \tag{A.9} \\
&\quad - \sum_\alpha \left( \delta U_\alpha \theta_\alpha + U_\alpha \left\lbrace \theta_\alpha , G \right\rbrace \right) \nonumber \\
& \qquad - \sum_{A_{i}} \left( \delta V_{A_{i}}\phi_{A_{i}} + V_{A_{i}} \left\lbrace \phi_{A_{i}} , G \right\rbrace \right) . \nonumber
\end{align}
But now into eq. (A.9) we can substitute
\begin{align}
\delta p_i\dot{q}^i +p_i &\delta \dot{q}^i = -\frac{\partial G}{\partial q^i}\dot{q}^i + \frac{d}{dt}\left( p_i \frac{\partial G}{\partial p_i}\right) - \dot{p}^i
\frac{\partial G}{\partial p_i}  \tag{A.10} \\
& = \frac{d}{dt}\left( p_i \frac{\partial G}{\partial p_i}-G\right) + \left[
\left( \frac{\partial}{\partial t} + \dot{U}_\alpha \frac{\partial }{\partial U_\alpha} + \dot{V}_{A_{i}} \frac{\partial}{\partial V_{A_{i}}}\right) \mu_{B_{j}} \right]\phi_{B_{j}}\nonumber
\end{align}
yielding
\begin{align}
\delta S_E = \int dt & \bigg[ \left( \frac{D}{Dt}\mu_{B_{i}}\right) \phi_{B_{i}} + 
 \mu_{A_{i}} \left( D_{A_{i}}^{B_{j}} \phi_{B_{j}} + D_{A_{i}}^\gamma \theta_\gamma \right) \tag{A.11} \\
&- \sum_\alpha \left( \delta U_\alpha \theta_\alpha - U_\alpha \mu_{B_{j}} \left(
C_{B_{j\alpha}}^\gamma \theta_\gamma + C_{B_{j\alpha}}^C \phi_C\right)\right)\nonumber\\
& - \sum_{A_{i}} \left( \delta V_{A_{i}} \phi_{A_{i}} - V_{A_{i}} \mu_{B_{j}}
 \left( C_{B_{j}A_{i}}^{C_{k}} \phi_{C_{k}} + C_{B_{j}A_{i}}^\gamma \theta_\gamma \right)\right) \bigg] .\nonumber
\end{align}
In eq. (A.11), we have dropped all surface terms, defined
\begin{equation}
\frac{D}{Dt} = \frac{\partial}{\partial t} + \dot{U}_\alpha \frac{\partial}{\partial U_\alpha} + \dot{V}_{A_{i}} \frac{\partial}{\partial V_{A_i}}\tag{A.12}
\end{equation}
and have used the fact the $\phi_{A_{i}}$ are all first class so that 
\begin{equation}
\left\lbrace \phi_{A_{i}} , H_C \right\rbrace = D_{A_{i}}^{B_{j}} \phi_{B_{j}}+ D_{A_{i}}^\gamma
\theta_\gamma . \tag{A.13}
\end{equation}
In eq. (11), we can arrange for $\delta S_E = 0$ by choosing $\delta U_\alpha$ so that all coefficients of $\theta_\alpha$ vanish, and by having the $\mu_{B_{i}}$ satisfy a differential equation that answers that the coefficients of $\phi_{B_{i}}$ sum to zero.  Upon having [19, 23  ] $\delta V_{A_{i}} = V_{A_{i}} = 0 ( i \geq 2)$, $S_E$ reduces to $S_T$, the total action, and $G$ becomes the generator of gauge transformations that leave
\begin{equation}
S_C = \int dt L(q^i, \dot{q}^i ) \tag{A.14}
\end{equation}
invariant, as $S_T$ and $S_C$ have the same dynamical content.

We can replace eq. (A.8) with
\begin{equation}
\delta f = \left\lbrace f, G \right\rbrace^*\tag{A.15}
\end{equation}
as by eq. (A.3), $\left\lbrace f, G \right\rbrace^*$ and $\left\lbrace f, G \right\rbrace$ differ by an expression that is at least linear in $\theta_\alpha$; in eq. (A.11) this term can be absorbed into $\delta U_\alpha$.  The advantage of using the DB over the PB in finding $\delta f$ is that we can set $\theta_\alpha = 0$ at the outset of any calculation.

It would be interesting to see how the approach of ref. [27] to finding gauge symmetries could be adapted to the case in which primary second class constraints are present.

\end{document}